# Driving and charging an EV in Australia: A real-world analysis

Thara Philip[1], Kai Li Lim[1], Jake Whitehead[1,2]
[1]The University of Queensland
[2] Electric Vehicle Council of Australia
Email for correspondence: t.philip@uq.edu.au

## Abstract

As outlined by the Intergovernmental Panel on Climate Change, electric vehicles offer the greatest decarbonisation potential for land transport, in addition to other benefits, including reduced fuel and maintenance costs, improved air quality, reduced noise pollution, and improved national fuel security. Owing to these benefits, governments worldwide are planning and rolling out EV-favourable policies, and major car manufacturers are committing to fully electrifying their offerings over the coming decades. With the number of EVs on the roads expected to increase, it is imperative to understand the effect of EVs on transport and energy systems. While unmanaged charging of EVs could potentially add stress to the electricity grid, managed charging of EVs could be beneficial to the grid in terms of improved demand-supply management and improved integration of renewable energy sources into the grid, as well as offer other ancillary services. To assess the impact of EVs on the electricity grid and their potential use as batteries-on-wheels through smart charging capabilities, decision-makers need to understand how current EV owners drive and charge their vehicles. As such, an emerging area of research focuses on understanding these behaviours. Some studies have used stated preference surveys of non-EV owners or data collected from EV trials to estimate EV driving and charging patterns. Other studies have tried to decipher EV owners' behaviour based on data collected from national surveys or as reported by EV owners. This study aims to fill this gap in the literature by collecting data on real-world driving and charging patterns of 239 EVs across Australia. To this effect, data collection from current EV owners via an application programming interface platform began in November 2021 and is currently live.

## 1. Introduction

Demand for transport is expected to grow significantly in the coming decades, and without significant initiatives to shift to the least carbon-intensive travel options, the carbon dioxide emissions from the transport sector could increase by 60 percent by 2050 (International Transport Forum, 2021). Electric vehicles (EVs) offer lower emissions compared to internal combustion engine vehicles (ICEVs) (European Commission, 2020). These emissions can be further reduced if EVs can be charged using 100 percent renewable energy.

Australian EV uptake rates are concerningly low at 2 percent of all car sales compared to adoption approaching 10 percent internationally (International Energy Agency, 2022). However, with EVs attaining competitive prices with ICEVs in the near future globally, if Australia can introduce policies that support increased availability of EV models locally, uptake could significantly improve. In a 2021 survey of over 3000 participants from across Australia, 51 percent of participants owned conventional vehicles and were considering switching to an EV for their next car purchase. Half of the respondents indicated a willingness to pay higher amounts for EVs than an equivalent conventional vehicle, and 55 percent indicated that they would use solar panels or battery storage to power their EVs (Electric Vehicle Council & carsales, 2021).

The extensive penetration of EVs in the market presents challenges and opportunities for power systems. On the one hand, EVs will increase energy demand. Global electricity demand for EVs was 80 TWh in 2019. EVs are projected to account for anywhere between 2 to 4 percent of total electricity demand in 2030, compared to 0.3 percent today (International Energy Agency, 2020). If EV charging events tend to overlap with current electricity demand peaks, it could potentially add additional stress to the grid. On the other hand, EVs could act as batteries-on-wheels, thereby storing excess energy during off-peak periods, when renewable energy generation peaks, and potentially transmitting energy back into the grid during peak electricity demand periods. The opportunities of EVs acting as batteries-on-wheels can be enabled through smart charging programs such as managed charging (V1G) and bi-directional charging, also known as vehicle-to-grid (V2G).

Total electricity demand from EVs will be determined by two factors: EV uptake and charging patterns. With EV uptake increasing, an emerging area of research is understanding EV owners' characteristics and usage patterns. To support grid stability and plan for future infrastructure upgrades, it is essential to predict the electricity demand by EVs in the future accurately. Therefore, understanding how and when EV owners drive and charge their vehicles is essential to assess and manage the impact of EVs on both transport and electricity systems. EV owners' decision to charge could be dependent on various factors, including owner characteristics and external factors. Examples of owner characteristics could be battery range, home charging facility, battery storage, rooftop solar, etc. External factors could be public fast-charging options, electricity tariffs, government policies, etc.

Currently, limited studies are exploring the driving and charging behaviour of EV owners. Some studies in this space have tried to estimate EV driving and charging behaviour based on stated preference surveys. Others have tried to elicit EV owners' behaviour using EV trial data. However, there is still limited available data on real-world EV driving and charging patterns (Lim, Speidel, & Bräunl, 2022; Weldon, Morrissey, Brady, & O'Mahony, 2016).

Through this study, 227 EV owners with 239 electric vehicles were recruited from across Australia (in addition to a further 163 EV owners internationally) through a partnership with API telematics platform Teslascope. To the best of the authors' knowledge, this is the first study of this kind in Australia and one of the first at a global level. Figure 1 shows the heat map of driving events logged so far across Australia, highlighting the spread of data being collected across the country.

Teslascope logs various levels of driving and charging decisions made by owners, and this data is also made available to the EV owners. As part of the signup agreement for the research program, Teslascope users agreed for their driving and charging data to be shared with our research team in exchange for a free 12-month subscription on the platform. This data includes distance driven, location, energy consumed during a trip, charging duration, time of charging, charging location, battery state of charge (SoC) at start and end of charging event, location of charging, etc.

**Figure 1: Heatmap of driving events**

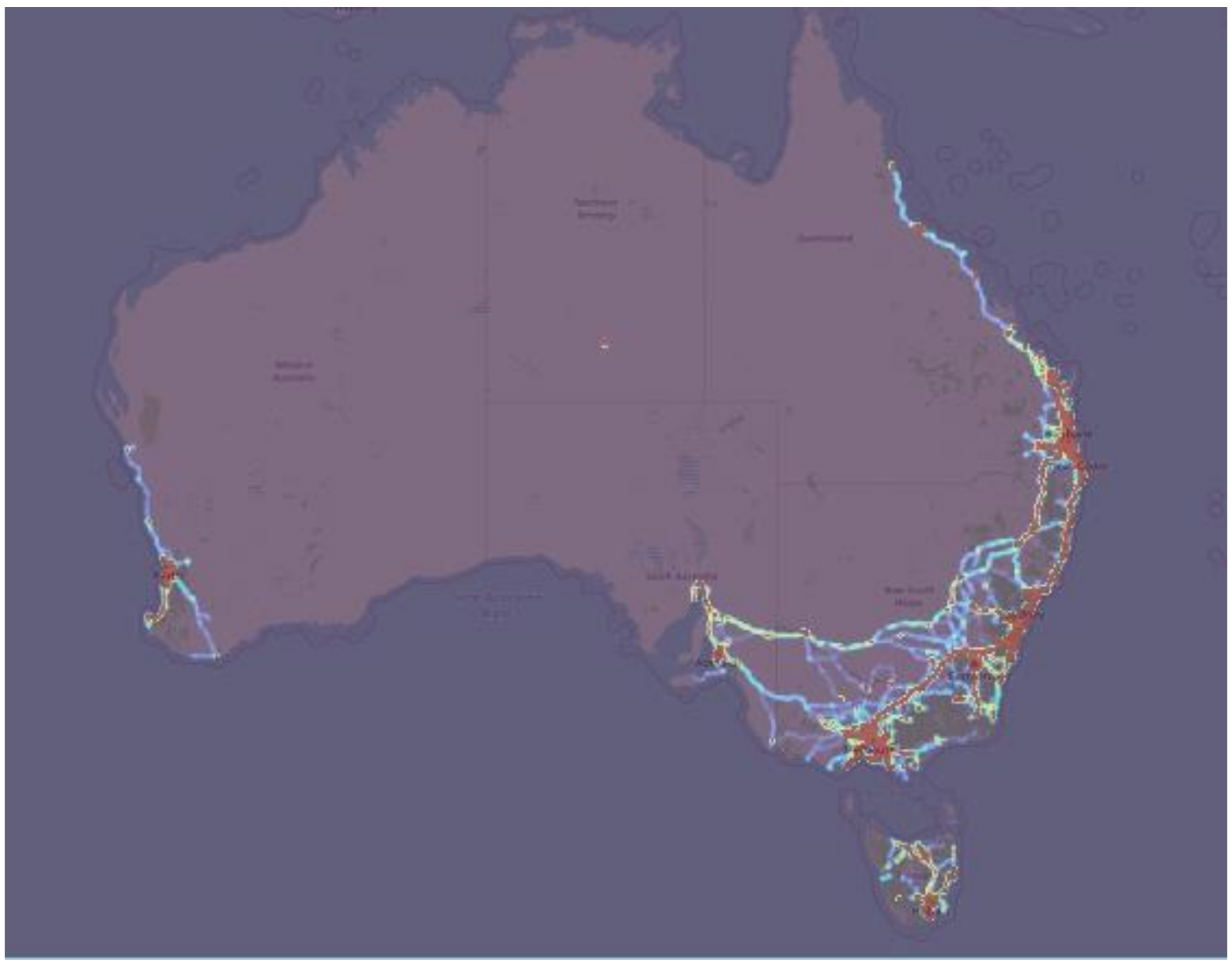

# 2. Background

Some of the existing studies in this space have used stated preference surveys to elicit EV owners' driving and charging behaviour (Ashkrof, de Almeida Correia, & van Arem, 2020; Daina, Sivakumar, & Polak, 2017; Jabeen, Olaru, Smith, Braunl, & Speidel, 2013; Lavieri & Oliviera, 2021; Yang, Yao, Yang, & Zhang, 2016). Stated preference surveys capture potential EV consumers' preferences under real and hypothetical scenarios. However, these may not reflect their actual behaviour.

There have been a few experimental studies investigating EV owner behaviour (Crozier, Morstyn, & McCulloch, 2021; Franke & Krems, 2013; Khoo, Wang, Paevere, & Higgins, 2014; Kim, Yang, Rasouli, & Timmermans, 2017; Lee, Chakraborty, Hardman, & Tal, 2020; Morrissey, Weldon, & O'Mahony, 2016; Sun, Yamamoto, & Morikawa, 2016; Weldon et al., 2016; Zoepf, MacKenzie, Keith, & Chernicoff, 2013). Some of these experimental studies relied on owner logbooks for data, while others relied on owners logging data for a short period, such as seven days, as used by Lee et al. (2020) and Crozier et al. (2021). Additionally, in some of these trials, the sample size has been small, e.g., only fifteen EVs were monitored in a study from Ireland (Weldon et al., 2016). Franke and Krems (2013) used car-based data loggers from BMW, in addition to EV owners' travel diary logs, to assess charging and driving behaviour. Khoo et al. (2014) analysed 4933 public and private charging events as part of the Victorian EV trial from 2010 to 2014. Under this program, 33 EVs were allotted to households and fleet participants for approximately three months. This study analysed 120 household participants and 70 fleet participants up to June 2013. Zoepf et al. (2013) estimated the probability of charging at the end of each trip by assessing the driving and charging behaviour of 125 prototype Toyota Prius Plug-in Hybrid Electric Vehicles (PHEVs) in the US deployed between April 2011 to April 2012.

It should be noted that the driving and charging patterns of PHEV owners may vary from that of BEV owners, given that PHEVs are not reliant on charging for propulsion. Some studies, such as Morrissey et al. (2016) have focussed on charging at public chargers and included a limited number of household charging events. Nevertheless, one of the common drawbacks of experimental studies is that the settings could influence the participants' usage patterns and not reflect real-world behaviour.

## 2.1. Driving choices of EV owners

In a 6-month study by Franke and Krems (2013) among 79 leased EVs in Berlin, it was reported that, on average, users drove 38 km per day. In a 2021 survey, 104 EV owners reported average daily driving distances twice the national average per car per year (Lavieri & Oliviera, 2021). However, the Australian Bureau of Statistics (2021) estimated that, on average, EV passenger vehicles in Australia in 2020 travelled the same distance as other passenger vehicles and 600 km further than petrol passenger vehicles (5.7 percent). Other studies have reported that EV owners in rural areas drive farther compared to their city counterparts. For example, drivers in a suburban territory in Canada were reported to drive 80 percent further than their urban counterparts on average (fleetcarma, 2019). This is not unexpected, with Khoo et al. (2014) reporting that the average driving distances of residents in rural areas or outer suburbs are likely to exceed the driving distances of residents in cities. Weldon et al. (2016) posited that EV owners in Ireland, on average, drove about 30 km between charging events, noting that the studied EVs had a relatively low driving range of 130 km.

## 2.2. Charging choices of EV owners

Lee et al. (2020) reported vehicle characteristics, commute behaviour, workplace charging availability, gender, and age as significant factors related to EV owners' charging location choices. Ashkrof et al. (2020) posited that EV owners' charging behaviour was affected by travel time, travel cost, SoC at origin and destination, and charging characteristics such as fast-charging duration and wait time at a fast-charging station, and access to a slow charging station at their destination. The majority of existing studies have identified heterogeneities in the charging behaviour of EV owners (Daina et al., 2017; Franke & Krems, 2013; Kim et al., 2017; Zoepf et al., 2013). An online survey in Australia of current and potential EV owners (Lavieri & Oliviera, 2021) reported that the home is the most preferred charging location and that residential charging could possibly peak during evenings if left unmanaged. Individuals who have rooftop solar panels could prefer to charge at home only (Jabeen et al., 2013). Lee et al. (2020) reported that more than half of Tesla owners exclusively rely on home charging. The study also posited that female owners preferred home over non-home charging locations and that non-home charging was necessary for owners in multi-unit dwellings. Multiple studies have identified that workplace charging or other car park locations were the second preferred location and public charging was the least prominent (GEOTAB Energy, 2020; Lee et al., 2020; Morrissey et al., 2016; Speidel & Bräunl, 2014). A study in Canada reported that although home charging was still the most preferred location, it is showing a decreasing trend which is attributed to increased workplace charging (fleetcarma, 2019).

In a survey involving 129 participants in New South Wales and the Australian Capital Territory from November 2019 to February 2020 (Ausgrid, 2020), the most popular time to charge was identified to be 10:00 pm to 7:00 am, with 53 percent almost always charging at this time. Weldon et al. (2016) reported a morning peak with the highest number of charging events starting between 7:30 am to 9:00 am. Contrary to expectations, a similar evening peak was not detected, with only a slight increase in the number of evening charging events. However, morning and evening peaks were reported during weekends. Morrissey et al. (2016) reported that EV owners in Ireland preferred to charge at home in the evenings and that incentivisation may encourage home charging at other times. Weldon et al. (2016) identified a morning charging peak. Lee et al. (2020) identified different charging patterns during weekdays and weekends. Khoo et al. (2014) reported negligible differences between weekday and weekend charging durations in Australia.

In a German study involving 79 participants who were using leased EVs, it was reported that EV owners charged their cars three times a week and that while many owners charged their EVs at higher SoC levels, a minority started charging at much lower SoC levels (Franke & Krems, 2013). The study also suggested that owners could possibly have a consistent battery charging style over time. In a survey involving 7,979 participants in the US, only 45 percent said that they charge their EVs daily, while the majority said that although they charge multiple times a week, it is not necessarily every day. Weldon et al. (2016) identified that most EV owners were charging in a regular pattern. A quarter of charging events started when the battery SoC was above 80 percent, potentially pointing to habitual or inefficient EV charging.

A US survey reported that the type of residence impacted EV owners' use of home charging and other secondary chargers. EV owners who lived in apartments or non-freestanding residences were likelier to charge in public or workplaces (GEOTAB Energy, 2020). Kim et al. (2017) posited that most EV owners charge their cars randomly at public charging stations while only a small number were regular users of a specific charging station.

## 2.3. Research gaps

It is notable that there are only limited studies investigating the actual driving and charging behaviour of EV owners, with the majority of existing literature relying on stated preferences or experiments. Additionally, some of the extant studies were based on deterministic rules. For instance, some studies assumed that EV owners would prefer to charge only if the SoC dropped below certain levels (Hu, Dong, & Lin, 2019; Yang et al., 2016). Another study assumed charging to be a function of variables such as SoC, time, and whether a journey has ended (Crozier et al., 2021). However, these could differ significantly from actual behaviour associated with EV charging as various other factors affect owners' decisions to charge and are not always deterministic, which could lead to overestimating the aggregated charging demand of EVs and unnecessarily upgrading electricity grids at a high cost. This highlights the importance of further studies into understanding the actual driving and charging behaviour of existing EV owners. One of the barriers to date to investigating these behaviours has been the high cost of collecting telematics data. Specialised data platforms will need to be developed to properly study this data across multiple sources and API formats in anticipation of vehicle manufacturers' increasing API connectivity features in the near future (Lim, Whitehead, Jia, & Zheng, 2021). This study employs an innovative approach to data collection using an API connection directly with EVs, removing the need for purchasing expensive telematics hardware.

## 2.4. Significance of research

This study aims to provide greater insight into Australian EV owners' driving and charging behaviour based on real-world telematics data. A significant contribution of this study is developing time-series data modelling to inform local EV patterns such as average daily driving distance, preferred times of charging, and power demand. Identifying consumers' preferences towards charging in terms of the time of day, duration, and preferred locations will support future work estimating the impact of EV charging on the electricity grid. The findings of this study will also help inform the planning of EV charging infrastructure and electricity grid infrastructure, as well as develop policies and energy tariff structures to support desirable EV charging behaviour. It could also potentially help reduce the adverse effects of various incorrect assumptions often made during the modelling of EV owners' behaviour for research and infrastructure development (Weldon et al., 2016).

Australia is well-placed geographically and spatially to take advantage of renewable energy generation, particularly solar, and is one of the countries with the highest rates of rooftop solar installation (Commonwealth Scientific and Industrial Research Organisation, 2021). Notably, however, Australia has low EV adoption rates compared to other developed countries with comparable national incomes. Switching to EVs is a critical aspect of decarbonising the Australian transport sector. Smart charging techniques, such as V1G and V2G, have been reported to positively impact consumers' EV preferences (Noel et al., 2019). The findings of this study will help to inform the extent to which behaviour changes will be required for the successful deployment of smart charging in Australia, if any, and in turn, inform what this might mean for future EV uptake. The findings of this study are also relevant for other emerging EV markets.

# 3. Methodology

Data on actual driving and charging behaviours of EV owners have been collected using a telematics platform called Teslascope. Tesla allows owners to grant application programming interface (API) access to their vehicles. This API access permits telematics data collection at highly competitive rates (e.g., less than USD 30 per vehicle per year) compared to traditional telematics, which depends on vehicle hardware installation (e.g., USD 500 per vehicle per year). Currently, Tesla is the only car manufacturer that widely allows API access. As and when other firms allow API access, this research can be extended to include other brands of cars as well. The participant recruitment started on 17 November 2021. Participants were incentivised with a one-year premium Teslascope membership worth USD 30. Each Teslascope participant can have multiple cars linked. Each of the different cars is identified by its Vehicle Identification Numbers (VIN).

# 4. Results and discussion

Data analysis was performed on the Teslascope dataset to understand Australian Tesla owners' driving and charging behaviour.

The comprehensive list of data measures is tabulated in Table 1, with data analysis presented across charging and driving events, as detailed in their respective subsections of this paper. This study uses time series analysis to discuss how charging and driving behaviours change across weekdays and weekends. In addition, spatial analysis is presented as heat maps for a case study on driving and charging locations in Brisbane, and econometric analysis is presented on the effects of demographic factors on EV owners' driving and charging behaviour.

**Table 1: Data analysis measures and overall statistics.**

| Measure | Statistic |
|---|---|
| **Sampling period start** | 2021-11-15 00:00 |
| **Sampling period end** | 2022-05-10 23:59 |
| **Sampling period** | 177 days |
| **Number of vehicles** | 239 |
| **Number of events** | 57,291 (driving); 19,575 (charging) |
| **Total distance driven** | 802,110.1 km |
| **Total on-road duration** | 1,388.49 days |
| **Total charging energy** | 248,383.4 kWh |
| **Total charging duration** | 2,334.49 days |

For analysis purposes and to augment data quality, the study interprets data according to the following filters:

- Driving events: Driving distance > 1 km
- Charging events: Charging energy consumption >1 kWh

It is also assumed that the EV is used by an individual throughout the study period, although it is likely to be used by their family or friends within the study period.

## 4.1. Charging data analysis

Charging events in the dataset are analysed as time and density plots to visualise charging demand. Time-series analyses are given as hour-of-day plots, and other plots analyse battery SoCs and energy consumption.

Figure 2a shows the density plot of the start of charging events by the hour of the day. During weekdays, it is observed that there is a morning charging peak from around 8:00 am to 12:00 pm. This pattern aligns with the findings of Weldon et al. (2016). Both weekends and weekdays are characterised by an increased number of charging events starting at 9:00 pm and a minor increase at midnight. This increase suggests EV owners are likely managing their charging times to take advantage of off-peak rates or Time-of-Use (ToU) tariff rates, coinciding with the start of off-peak tariffs from energy retailers (off-peak at 9:00 pm and super off-peak at midnight). Note that this charging behaviour is self-managed and not enforced by the grid operator or energy retailer.

Furthermore, Figure 2b shows active charging density (continuous plugged-in duration) across the hours of the day. It integrates into Figure 2a, where the gradient of the curves indicates the change in charging events. For example, a positive gradient indicates an increase in vehicles plugging in for charging and vice versa. Similarly, most EVs are plugged in at night—charging peaks around 10:00 pm to 1:00 am. However, most EVs are plugged-in at midday, which would indicate a combination of EVs charging at workplaces or off solar energy at home. Charging utilisation is lowest at 7:00 am and 7:00 pm, which coincides with morning and evening peak tariffs, as noted with the decrease in charging events from 4:00 pm onwards.

**Figure 2: Frequency of charging events plotted by (a) charging start time (b) charging utilisation (plugged-in) time**

**(a) Charging start** **(b) Charging utilisation**

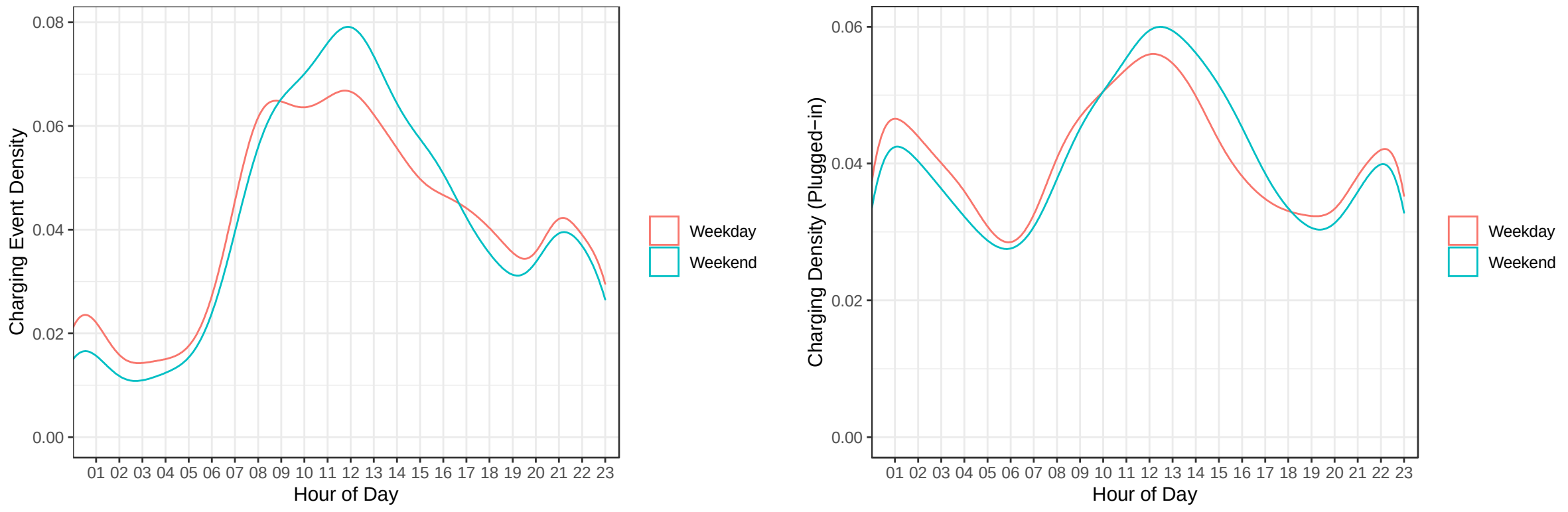

Combining the inferences from Figures 2b and 4b, Figure 3 aggregates the average power delivered to a charging EV at each hour of the day. The higher density of charging events shown in Figure 2 at noon contributes to average energy peaks of 368 W per vehicle and 423 W per vehicle on weekdays and weekends, respectively (including home and public charging). A smaller peak occurs at midnight when most EVs are likely charging using an off-peak tariff.

Conversely, low power consumption is observed in the morning and evening peak periods. Aggregation assumes a constant energy delivery across the entire charging duration for simplicity. The mean power delivery is 250 W hourly and 6 kWh daily across the entire dataset, including home and public charging. On average, this sample of EV owners has been observed to charge 0.463 times per day or around once every two days.

**Figure 3: Average power delivered to charge an EV for each hour of the day**

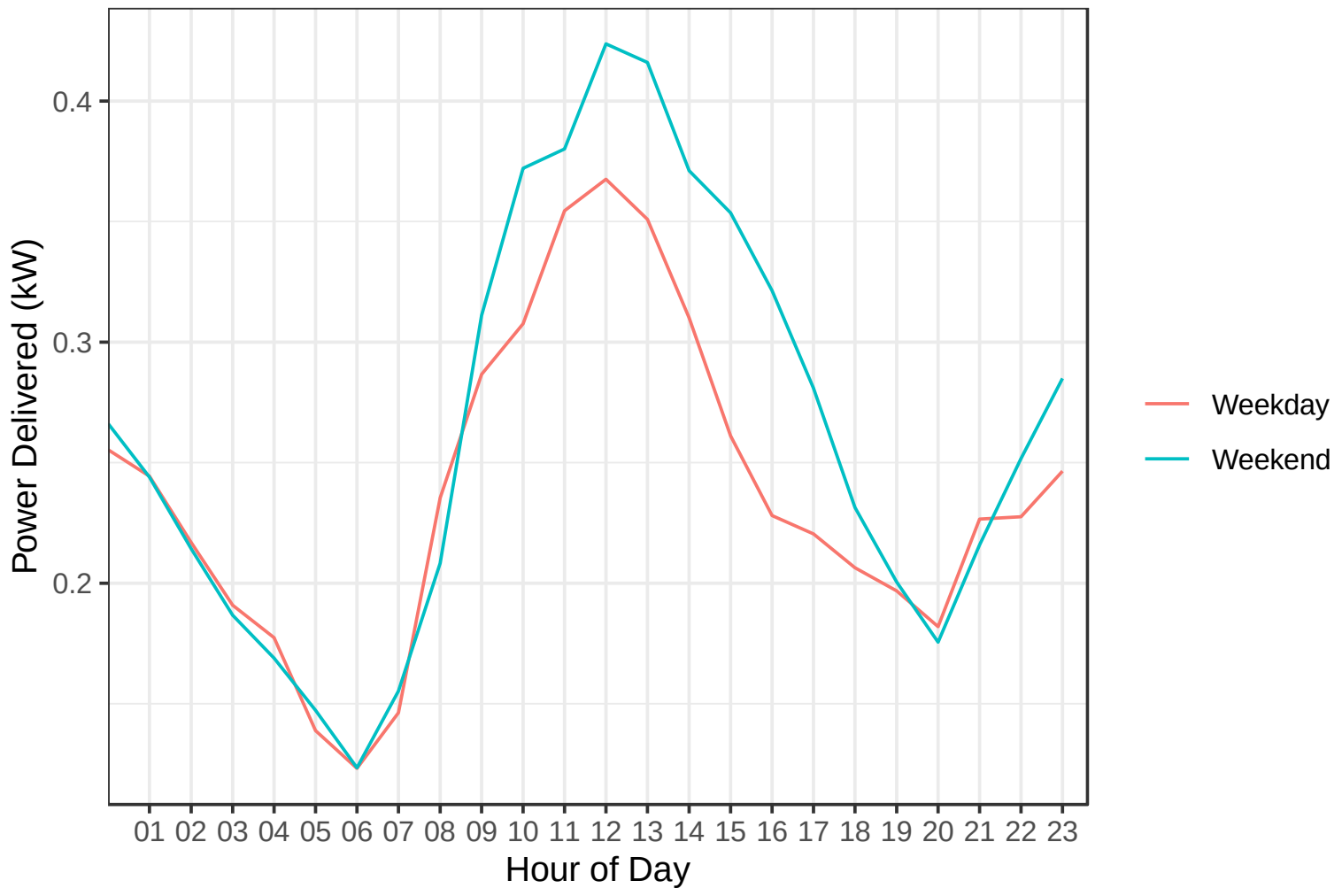

Figure 4a shows the density plot of charging events by the hour of the day. The majority of charging events start when battery SoC is between 50 to 80 percent, with a mean of 59.1 percent. Similar patterns are observed for both weekdays and weekends, with a slightly higher charging peak at 75 percent battery SoC during weekdays. This difference could denote a lower number of plug-ins during weekends. Overall, this behaviour indicates that most events are top-up charges, with deep charges (i.e., charging more than 80 percent battery capacity) rarely observed. Drawing further inferences, this behaviour could be interpreted as regular top-ups during the week, with longer trips during the weekends leading to deeper charges.

Aligning with the top-up charging behaviour observed in Figure 4a, Figure 4b illustrates that the charging energy consumption for most events is less than 15 kWh, i.e., less than 25 percent of the smallest Tesla vehicle's battery size (Model 3 at 60 kWh). The mean charging energy consumption is 12.7 kWh per charging event, with more frequent top-up behaviour observed on the weekdays.

**Figure 4: Frequency of charging events by (a) starting battery SoC and (b) energy delivery per charge**

**(a) Starting battery SoC** **(b) Energy delivered**

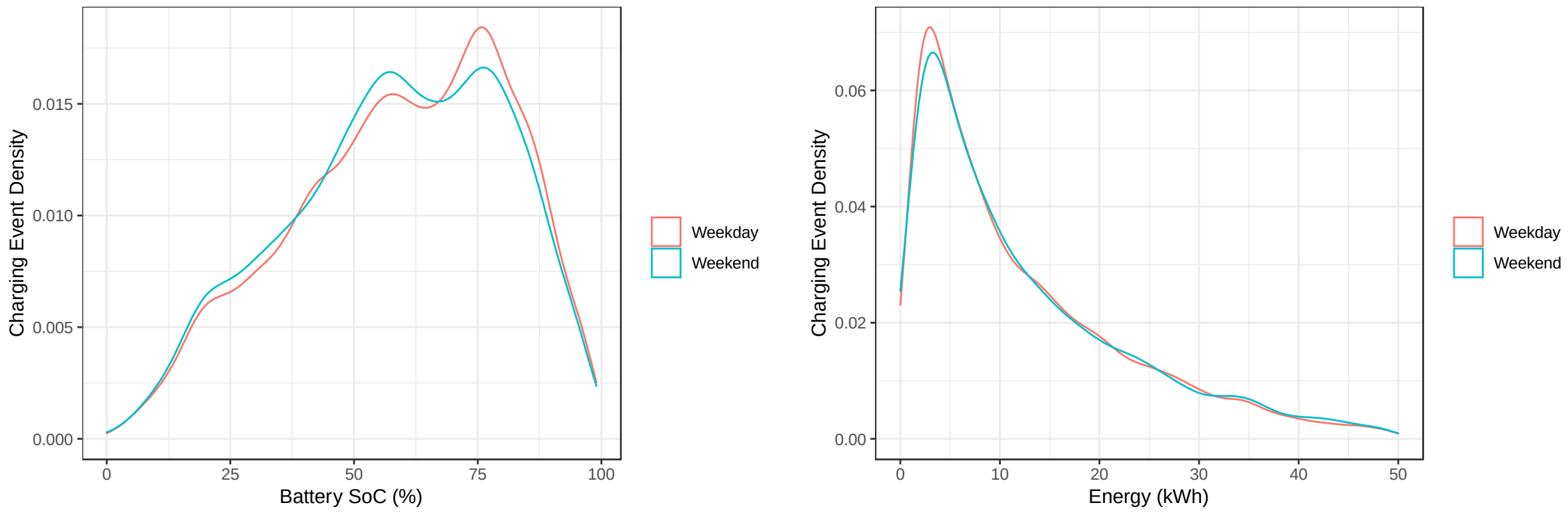

To summarise, as opposed to the popular claim that EVs present a significant risk to the grid, this data suggests in large part that EV owners are already successfully managing their charging to occur outside of peak periods, with existing tariffs appearing to send a strong enough price signal to induce this behaviour at present, combined with EV owners preferencing daytime charging to take advantage of home solar PV. Furthermore, with an average of a 219 W increase in power demand per EV in the 6 pm peak period across the entire grid, i.e., including both home and public charging (see Figure 3Figure 3), it is expected that the impact of EV charging on the grid is likely to be minimal for the foreseeable future.

Encouragingly, with charging behaviour primarily already aligning with ideal charging patterns that would be the objective of future smart charging programs, it is expected there are significant opportunities to shift EV owners to these programs as they are introduced to the market over the coming years with minimal changes in behaviour, and in turn, potentially minimal costs. Consequently, these findings highlight that it would be premature to introduce charging management regulations that could negatively affect EV adoption and jeopardise Australia's pathway to achieving net-zero carbon emissions. Instead, a more prudent approach would be to continue incentivising and encouraging the existing positive charging behaviour until when or if that behaviour significantly changes or for specific user groups that appear more likely to charge during the evening peak period (see Section 4.4).

## 4.2. Driving data analysis

Driving behaviour is analysed in terms of both events and distances. These are presented as density plots. Figure 5a shows the density of driving times (on-road) across each hour of the day. It is discerned that on weekdays, driving event occurrences have a morning and evening peak, as is expected during the workday commuting hours. They peak around 8:00 am in the morning and 4:00–5:00 pm in the evenings; intra-day driving is less frequent but with more produced peaks due to dynamic driving activities. The driving event starts to peak at 11:00 am on weekends, with no significant evening peak identified. This weekend pattern is more consistent with leisure or road trips that typically begin in the late morning.

Analysing driving distances in Figure 5b, it was noted that EVs have similar commuting patterns as most passenger vehicles in the country. Many EVs are driven less than 50 km per day, with a mean distance of 41.8 km, excluding days when the vehicle is not driven. With non-driving days aggregated, the average monthly driving distance per EV is 795.2 km. This distance is slightly lower than the Australian Bureau of Statistics findings in 2020 (Australian Bureau of Statistics, 2020), which could be due to a reduction in commutes as a consequence of the COVID-19 pandemic. Weekday drives are generally shorter, as they typically cover regular commutes. EVs often drive less than 10 km per driving event, with most drives averaging 8.5 km. These telematics findings differ from Lavieri and Oliviera (2021), which suggested that, on average, annual distances driven by EV owners were twice the national average, noting that this was based on survey responses as opposed to telematics data.

**Figure 5: Frequency of (a) on-road driving events by the hours of the day and (b) driving distances per day**

**(a) Driving events** **(b) Driving distances**

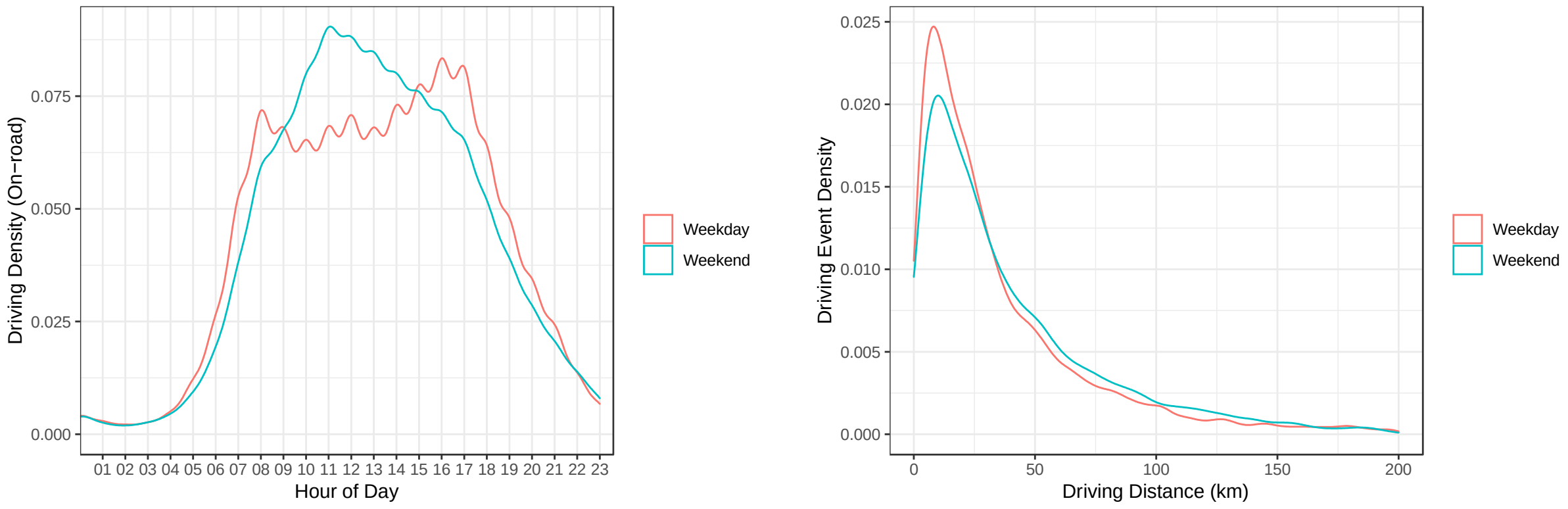

The analysis of the real-world driving and charging behaviour of EV users provides favourable insights into the potential of deploying smart charging programs in Australia. With the daily average driving distances of EVs relatively low, this means that EVs are parked for a significant period of the day, including during the evening peak electricity demand period (4:00–8:00 pm). In terms of charging behaviour, the two peak charging periods already appear well-aligned with renewable energy generation during the day (solar) and overnight (wind). These findings further strengthen the case for using EVs as batteries-on-wheels to absorb excess renewables and potentially export this back to the grid during peak electricity demand periods.

## 4.3. Heat maps: Brisbane case study

Spatial analysis of driving patterns is performed through heat map visualisations. Figure 8 shows the heat maps of charging and driving events around the Brisbane inner-city area as a case study. This region was selected due to its proximity to The University of Queensland for verification purposes. The heat maps annotate the density of driving and charging events, with the colour palette ranging from purple (minimum density) to maroon (maximum density), scaling between 0 and 200 events.

From the driving map in Figure 6a, it is observed that most traffic is concentrated along major state and arterial routes, along with significant movements in the Brisbane CBD and Fortitude Valley precincts. This result is consistent with Queensland’s traffic census data (Queensland Government, 2021), revealing that traffic is most prevalent on major highways and business districts.

Figure 6b shows the density heat map of charging locations around Brisbane. This heat map shows the most frequent locations for public DC fast-charging, with slower AC charging filtered from the dataset as they contain more private, home, and workplace charging. It was noted that the most popular public fast-charging stations are located at the Tesla Superchargers, which are in the Fortitude Valley and were previously in Toombul (before being closed due to the 2022 Brisbane floods), The University of Queensland (UQ St Lucia), Queensland University of Technology (QUT) Gardens Point, QUT Kelvin Grove, and others, as shown in Figure 6b.

**Figure 6: Heat maps showing (a) driving locations and (b) charging locations**

**(a) Driving locations** **(b) Charging locations**

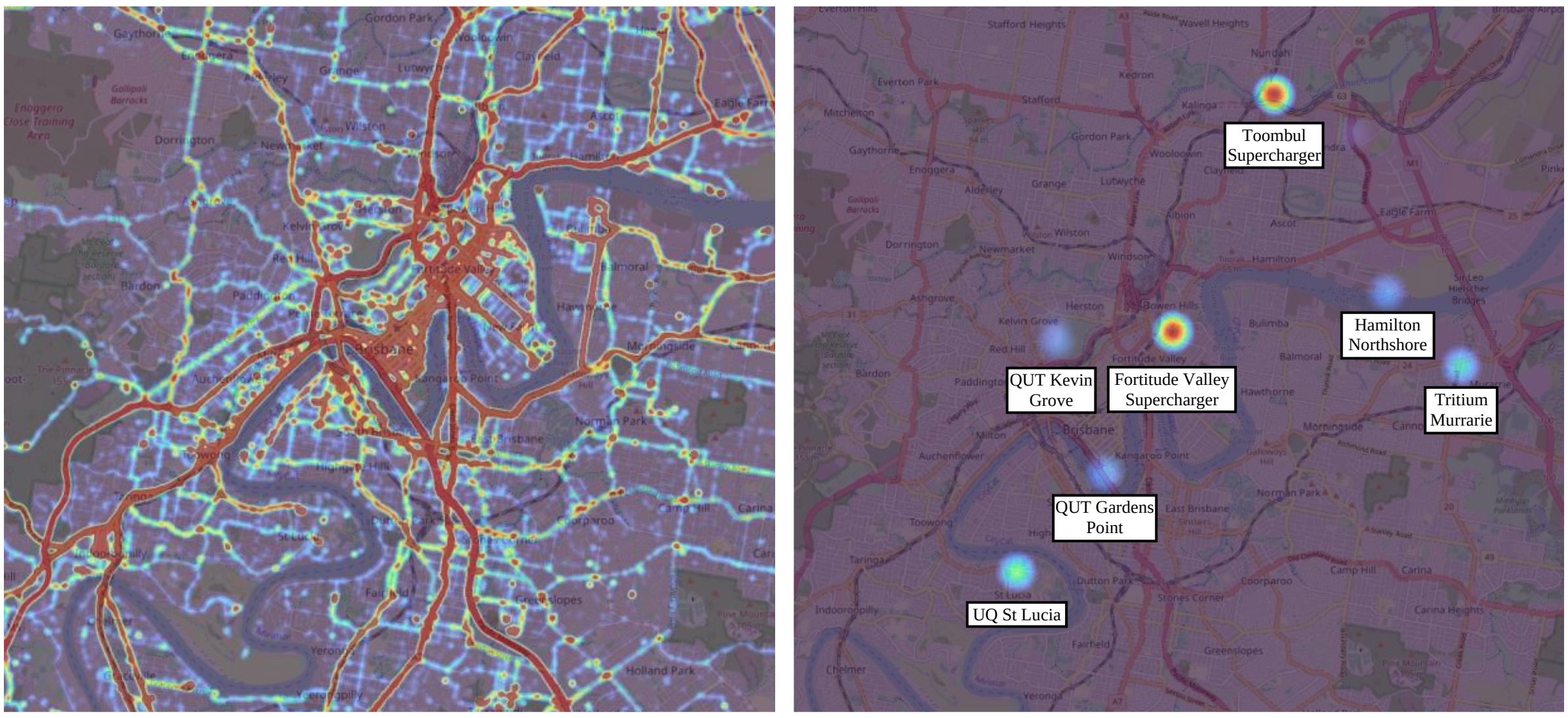

## 4.4. Econometric modelling

Two multinomial logit models were estimated to better understand some of the factors influencing EV owners' decision to charge during the evening peak electricity demand period and the decision to use fast-charging infrastructure.

### 4.4.1 Choosing to charge during the evening peak

Table 2 includes the model estimates for EV owners' decision to charge during the evening peak period (4:00–8:00 pm). It is observed that travelling long distances on the same day and the following day is positively associated with the choice to charge during the evening peak period. This could be a consequence of needing to charge over a more extended evening period to replenish the vehicle battery's SoC from the current day's trips or in preparation for the following day's trips.

It is also observed that higher odometer readings, owning a home (as opposed to renting), and having a home battery, are all factors that have a statistically significant relationship with EV owners' decision to charge during the evening peak period. It could be the case that higher odometer readings are acting as a proxy for EV owner experience and, in turn, owner awareness, particularly regarding the economic and electricity grid benefits of charging during off-peak periods.

EV owners that rent their residence may have less control over their home's electricity tariffs, removing the economic benefit of charging during off-peak periods. It could also be that renters are unable or less willing to install dedicated charging infrastructure at their home and are therefore reliant on slow, powerpoint-based mobile chargers, which provide less control over charging times and necessitate the need for longer charging duration due to the slower charging rate (in comparison to a hard-wired home EV charger). Notably, 89 percent of the EV owners in the sample reported owning their own homes.

Finally, EV owners with home batteries appear to be less likely to charge their EV during the evening peak period. This is likely due to these owners wanting to prioritise the energy stored in their home batteries to power the remainder of their household needs during this period to avoid paying for peak evening electricity from the grid. They are likely also on a time-of-use tariff, which would provide an economic benefit for charging their EVs during off-peak periods.

Further investigation is ongoing as part of this research project to identify and analyse the different electricity plans EV owners have selected and the influence of these plans on EV charging behaviour.

It is worth noting that no statically significant effects were established for gender, age, home type, access to home solar, or house location—suggesting these factors may be less significant in terms of which EV owners choose to charge during the evening peak period.

**Table 2: Multinomial logit model estimates for the decision to charge during the evening peak period. Headers denote the beta coefficient, standard error, p-value, and 95 percent confidence interval, respectively**

| Dependent Variable: Decision to charge during evening peak period | *β* | SE | *P* | 95 % CI | |
|---|---|---|---|---|---|
| **Constant** | -1.044 | 0.334 | 0.002 | -1.699 | -0.390 |
| **Charge energy added (kWh)** | -0.001 | 0.004 | 0.759 | -0.009 | 0.007 |
| **Odometer reading (km)** | 0.000** | 0.000 | 0.021 | 0.000 | 0.000 |
| **Distance travelled same day (km)** | 0.008*** | 0.001 | 0.000 | 0.007 | 0.009 |
| **Distance travelled next day (km)** | 0.001** | 0.000 | 0.027 | 0.000 | 0.002 |
| **Own home** ***(baseline: rent home)*** | -0.523** | 0.247 | 0.034 | -1.008 | -0.039 |
| **Live in a freestanding house** ***(baseline: Live in apartment/townhouse)*** | -0.049 | 0.217 | 0.823 | -0.475 | 0.377 |
| **Have rooftop solar** ***(baseline: do not have rooftop solar)*** | -0.142 | 0.225 | 0.528 | -0.583 | 0.299 |
| **Have home battery** ***(baseline: do not have a home battery)*** | -0.439** | 0.198 | 0.026 | -0.826 | -0.051 |
| **Owners 45 years or older** ***(baseline: Owners less than 45 years old)*** | -0.175 | 0.159 | 0.272 | -0.486 | 0.137 |
| **Male EV owner** ***(baseline: Female EV owner)*** | 0.337 | 0.303 | 0.266 | -0.257 | 0.932 |
| **Live in a city** ***(baseline: Live in a region)*** | -0.123 | 0.161 | 0.446 | -0.438 | 0.193 |
| **Model diagnostics** | | | | | |
| **Number of observations** | 11,627 | | | | |
| **Number of vehicles** | 239 | | | | |
| **Log pseudolikelihood** | -5760.305 | | | | |
| **Wald $\chi^2(11)$** | 225.99 | | | | |
| **Prob > $\chi^2$** | 0 | | | | |

*    statistical significance at $P < 0.10$ level
**  statistical significance at $P < 0.05$ level
***statistical significance at $P < 0.01$ level

### 4.4.2 Choosing to use fast charging infrastructure

Table 3 includes the multinomial logit model estimates of EV owners' decision to use fast-charging infrastructure. It is observed that amount of charge energy added on the same day, distance travelled the same day, distance travelled the next day, and living in a standalone house are all factors significantly associated with the decision to use fast-charging infrastructure.

As expected, EV owners who travel longer distances are more likely to use fast-charging infrastructure—likely to fulfil their travel requirements or as a top-up during a rest break mid-trip. Although it is to a lower extent, the use of fast-charging infrastructure is also observed to be positively associated with travelling longer distances the following day. This could reflect owners relying on fast-charging infrastructure to ensure they have maximum SoC in preparation for an upcoming long-distance trip.

Finally, the estimates suggest that EV owners who live in freestanding houses are less likely to use fast-charging infrastructure, and conversely, those living in apartments or townhouses are more likely to use fast-charging infrastructure. This finding is again within expectations and provides additional evidence to support the notion that fast-charging infrastructure is essential for long-distance trips and EV owners with limited access to charging infrastructure at their homes.

No statically significant effects were established for gender, age, home ownership, access to a home solar or home battery, or house location (city/region)—suggesting these factors may be less significant in terms of which EV owners choose to use fast-charging infrastructure.

**Table 3: Multinomial logit model estimates for the decision to use fast-charging infrastructure. Headers denote the beta coefficient, standard error, p-value, and 95 percent confidence interval, respectively**

| Dependent Variable: Decision to use fast charging infrastructure | *β* | SE | *P* | 95 % CI | |
|---|---|---|---|---|---|
| **Constant** | -3.977 | 0.652 | 0.000 | -5.254 | -2.700 |
| **Charge energy added (kWh)** | 0.073*** | 0.007 | 0.000 | 0.060 | 0.086 |
| **Odometer reading (km)** | 0.000 | 0.000 | 0.338 | 0.000 | 0.000 |
| **Distance travelled same day (km)** | 0.021*** | 0.001 | 0.000 | 0.019 | 0.023 |
| **Distance travelled next day (km)** | 0.002*** | 0.001 | 0.005 | 0.001 | 0.004 |
| **Own home** ***(baseline: rent home)*** | -0.109 | 0.413 | 0.793 | -0.917 | 0.700 |
| **Live in a freestanding house** ***(baseline: Live in apartment/townhouse)*** | -0.734** | 0.351 | 0.036 | -1.421 | -0.047 |
| **Have rooftop solar** ***(baseline: do not have rooftop solar)*** | -0.135 | 0.333 | 0.685 | -0.788 | 0.517 |
| **Have home battery** ***(baseline: do not have a home battery)*** | 0.126 | 0.312 | 0.686 | -0.485 | 0.737 |
| **Owners 45 years or older** ***(baseline: Owners less than 45 years old)*** | -0.173 | 0.243 | 0.476 | -0.649 | 0.303 |
| **Male EV owner** ***(baseline: Female EV owner)*** | -0.547 | 0.670 | 0.414 | -1.860 | 0.766 |
| **Live in a city** ***(baseline: Live in a region)*** | 0.242 | 0.274 | 0.378 | -0.295 | 0.778 |
| **Model diagnostics** | | | | | |
| **Number of observations** | 11,627 | | | | |
| **Number of vehicles** | 239 | | | | |
| **Log pseudolikelihood** | -2629.41 | | | | |
| **Wald $\chi^2$(11)** | 655.34 | | | | |
| **Prob > $\chi^2$** | 0 | | | | |

* statistical significance at $P < 0.10$ level
** statistical significance at $P < 0.05$ level
***statistical significance at $P < 0.01$ level

# 5. Conclusion

This paper has presented the first insights from The University of Queensland's Teslascope trial in its sixth month. This study intends to better inform future Australian EV-related decisions, including policy and regulation. The analysis described in this paper provides a unique perspective into Australian EV owners' charging and driving behaviour, finding that most EVs are used for daily commuting and leisure trips. When considering the behavioural differences across weekdays and weekends, they are comparable to general traffic expectations. Charging behaviour during weekdays appears to be mainly in the form of top-up charging events, compared to marginally longer charging durations during weekends.

Importantly, this study highlights that many EV owners are already taking advantage of off-peak electricity tariffs and solar power to charge their EVs. During the evening peak, EVs appear only to be increasing power demand by an average of 250 W per vehicle (home and public charging combined). This is minimal compared to increases caused by other electrical appliances during the same period and suggests that fears of EVs placing a significant burden on the electricity grid in the short term are likely overplayed and undoubtedly premature. However, this does not remove the need to plan future smart charging programs that support and encourage these existing, positive charging behaviours.

Looking to the future, it will be imperative to support renters in being able to shift charging events away from the evening peak period through greater access to off-peak tariffs and dedicated home charging infrastructure. In addition, increasing awareness of the economic benefits of charging during off-peak periods may also help to shift the minority of charging events occurring during the evening peak electricity period.

# Acknowledgements

The authors would like to thank all Teslascope trial participants for contributing to the dataset. This research is co-funded by an Advance Queensland Industry Research Fellowship and iMOVE, supported by the Cooperative Research Centres program, an Australian Government initiative. The authors also thank Dr Andrea La Nauze from the University of Queensland School of Economics for valuable comments on the econometric models.